# Dual-band, double-negative, polarization-independent metamaterial for the visible spectrum

Muhammad I. Aslam[a,b] and Durdu Ö. Güney[a, 1]

[a] *Department of Electrical and Computer Engineering, Michigan Technological University, Houghton, Michigan 49931, USA*

[b] *Department of Electronic Engineering, NED University of Engineering and Technology, Karachi 75270, Pakistan*

**Abstract**

We present the first dual-band negative index metamaterial that operates in the visible spectrum. The optimized four-functional-layer metamaterial structure exhibits the first double-negative (i.e., simultaneously negative permittivity and permeability) band in the red region of the visible spectrum with a figure of merit of 1.7 and the second double-negative band in the green region of the visible spectrum with a figure of merit of 3.2. The optical behavior of the proposed structure is independent of the polarization of the incident field. This low-loss metamaterial structure can be treated as a modified version of a fishnet metamaterial structure with an additional metal layer of different thickness in a single functional layer. The additional metal layer extends the diluted plasma frequency deep into the visible spectrum above the second order magnetic resonance of the structure, hence provides a dual band operation with simultaneously negative effective permittivity and permeability. Broadband metamaterials with multiple negative index bands may be possible with the same technique by employing higher order magnetic resonances. The structure can be fabricated with standard microfabrication techniques that have been used to fabricate fishnet metamaterial structures.

---

[1] Corresponding author e-mail address: dguney@mtu.edu

**Introduction**

Studies on negative index electromagnetic materials date back to the early 1900s [1]. For example, A. Schuster discussed the backward electromagnetic waves and their implications for optical refraction in 1904 [2]. He cited that the backward electromagnetic waves can propagate in sodium vapor within its absorption band. In 1905, H. C. Pocklington studied wave propagation in backward wave medium with a negative group velocity [3]. This work was revisited by G. D. Malyuzhinets in 1951 [4]. D. V. Sivukhin discussed general properties of wave propagation in negative index medium in 1957 [5].

In 1968, Veselago systematically showed that unusual and interesting properties such as negative refraction, reverse Doppler's effect, and reverse Cherenkov effect can be achieved using the materials having simultaneously negative permittivity and permeability [6]. In contrast to the right-hand rule followed in the naturally occurring dielectric materials, the electric and magnetic fields in such materials follow left-hand rule. Thus, these materials are named as left-handed materials (LHM). Due to the unavailability of LHM in nature, the negative index medium did not gain much attention until Pendry, et al. proposed structures to artificially achieve negative permittivity [7, 8] and negative permeability [9]. Following these findings, Smith, et al.

experimentally demonstrated LHM [10, 11] using periodic arrangement of split-ring resonators (SRR) providing negative permeability and thin metallic wires providing negative permittivity. These findings attracted significant amount of research attention to the design and application of metamaterials (materials having unusual optical properties) resulting in novel metamaterials at different frequency regimes [12-14]. The effective permittivity ($\varepsilon = \varepsilon' + j\varepsilon''$) and permeability ($\mu = \mu' + j\mu''$) of metamaterials can be controlled by varying their underlying constituents and/or geometry [15]. The research in the area of metamaterials is motivated by the wide variety of their potential applications such as, high precision lithography [16], perfect lens [17], high resolution imaging [17, 18], invisibility cloaks [19], small antennas [20], optical analog simulators [21, 22], and quantum levitation [23].

The traditional SRR based metamaterials cannot be scaled to optical and visible frequencies [24]. However, metamaterials based on the interaction of surface plasmon polaritons (SPP) are good candidates to achieve the negative refractive index ($n = n' + jn''$) at infrared (IR) and visible frequencies [14, 25-28]. The SPPs are the surface waves generated by the interaction of electromagnetic waves with an interface of two materials with different signs of permitivity and/or permeability [29-31]. Perhaps, the most important metamaterial proposed for optical frequencies is the fishnet metamaterial structure consisting of air holes drilled in the alternating layers of metal and dielectrics [32, 33]. Most of the fishnet metamaterials reported so far are functional for one specific polarization of the incident field and exhibit single negative index band. However, some researchers have proposed polarization independent fishnet metamaterials [34-38]. It has also been shown that fishnet metamaterials can support multiple magnetic resonances corresponding to the excitation of different orders of SPP modes [27]; however, only up to two negative index bands (i.e., dual-band) have been demonstrated in the sub-THz [38], near-IR [39], and near-IR/visible [40] regimes. In particular, Refs. 39 and 40 determine the state-of-the-art and limits of the dual-band negative index metamaterial structures. In Ref. 39, two fishnet magnetic resonators of different lateral dimensions are used in a single functional layer to achieve a dual-band operation. However, this complicates the multilayer fabrication. On the other hand, in Ref. 40, a single functional layer of metamaterial structure with a dual-band operation is experimentally demonstrated at near-IR and red regions. Although the structure provides double-negative (i.e., low loss) operation at near-IR, it works only as a single-negative (i.e., very lossy) metamaterial in the red region. Furthermore, this structure is not a dual-band metamaterial for a single polarization, since each negative index band requires a different orthogonal polarization. Therefore, different techniques are necessary for further frequency scalability of truly dual-band polarization independent negative index metamaterials operating in the visible spectrum. In contrast, our modified fishnet structure provides negative permittivity deep inside the visible spectrum. We also employ the higher order magnetic resonance of the structure to achieve two polarization independent negative index bands in the visible spectrum both with simultaneously negative permittivity and permeability (i.e., double-negative). To date, a single-band double-negative (DNG) metamaterial has been experimentally demonstrated to operate only up to the red region of the visible spectrum using fishnet metamaterials [34, 40-42] with a figure of merit $(\text{FOM} = -n'/n'')$ of 3.34 at 408THz [34]. We have theoretically reported the shortest operating wavelength for a single band negative index metamaterial at the violet region of the visible spectrum based on surface plasmon polaritons of thin metal films [25].

In this report, we have modified the traditional fishnet metamaterial by adding an additional metal layer in a functional layer. For low-loss negative index operation, the magnetic resonance frequencies should be below but close to the diluted plasma frequency. The additional metal layer in our structure increases the diluted plasma frequency slightly above the second order magnetic resonance of the fishnet structure to achieve a dual-band low-loss negative index band in the visible spectrum. The optimized four-functional-layer structure exhibits the first double-negative band in the red region of the visible spectrum with a figure of merit of 1.7 and the second double-negative band in the green region of the visible spectrum with a figure of merit of 3.2. The two-dimensional symmetry of the structure and the subwavelength features ensure the polarization independent behavior for the normally incident light. This is the first dual-band negative index metamaterial in the visible spectrum that has been reported.

**The modified-fishnet metamaterial**

Our modified fishnet metamaterial is shown in Fig. 1, with the coordinate axes and the polarization configuration of the normally incident wave. Due to the symmetry, the modified fishnet structure has the same behavior for different lateral polarizations of the normally incident field. This polarization-independent behavior is also numerically verified by simulating the structure with different angles of polarizations of the incident field. The structure consists of a stack of alternating layers of metal (silver) and dielectric (Magnesium Fluoride). Thickness of each dielectric layer is $t_i$ = 6nm and thicknesses of alternating silver layers are $t_{m1}$ = 20nm and $t_{m2}$ = 10nm. One functional layer of the modified fishnet metamaterial consists of two metal layers and two dielectric layers with a total thickness of 42nm. In order to achieve the polarization independence, a square lattice is used in the lateral plane with a lattice constant $p$ = 240nm. The side length of the sub-wavelength square aperture is 145nm. All simulation parameters are summarized in Table 1.

Frequency domain analysis of commercially available CST Microwave Studio software is used to calculate the *s*-parameters corresponding to the reflection and transmission coefficients at the input and output ports of the structure. These *s*-parameters are then used to retrieve effective parameters for the metamaterial. For numerical purposes, silver is described by the Drude model with the bulk plasma frequency of $f_p = 2180 \text{THz}$ and the collision frequency $\gamma_e = 13.5 \text{THz}$ [43]. In general, during fabrication, an extra dielectric layer is deposited on both sides of the structure to prevent the metal from the environmental effects [34, 37]. This layer is also used in the simulation (as shown in Fig. 1).

In order to calculate the s-parameters, ports were used at the top and bottom boundaries of the computational domain (see Fig. 1). Both ports were defined 30nm away from the metamaterial boundaries. A built-in phase-compensation technique in CST was used to evaluate the s-parameters at the metamaterial boundaries. We verified that further port distances have negligible effect on the results. Other boundaries were simulated as perfect electric conductor (PEC) and perfect magnetic conductor (PMC). From the computational point of view, PEC-PMC boundaries correctly simulate the periodicity of this structure with less memory and CPU time compared to the periodic boundary conditions. Final results are verified by simulating the structure with periodic boundary conditions. A built-in auto-meshing algorithm of CST was used (with adaptive meshing) to generate tetrahedral meshes. The transmission and reflection spectra for the modified fishnet metamaterial obtained using CST simulation are shown in Fig. 2. There are two extraordinary transmission (EOT) bands corresponding to the two negative index bands of the metamaterial.

**Effective behavior of the modified-fishnet metamaterial**

A metamaterial can be represented as an effective medium (with the same s-parameters) if its dimensions are sufficiently small compared to the operating wavelength. The s-parameters are related to the material parameters ($n$ and $z = z' + jz''$; where $z$ is the impedance) as [44-46]:

$$s_{11} = -\frac{j}{2}\left(z - \frac{1}{z}\right)\sin(nkd) \qquad (1)$$

$$\frac{1}{s_{21}} = \cos(nkd) - \frac{j}{2}\left(z + \frac{1}{z}\right)\sin(nkd) \qquad (2)$$

where $k$ is the free space propagation constant and $d$ is the length of the material in the propagation direction. The effective material parameters using isotropic retrieval are obtained by inverting equations (1) and (2), which gives

$$z = \pm\sqrt{\frac{(1+s_{11})^2 - s_{21}^2}{(1-s_{11})^2 - s_{21}^2}} \qquad (3)$$

$$n = \pm\frac{1}{kd}\cos^{-1}\left(\frac{1 - s_{11}^2 + s_{21}^2}{2 s_{21}}\right) \qquad (4)$$

The sign ambiguity is resolved by causality and passivity of the metamaterial that respectively require $z' > 0$, $n'' > 0$ [44-46]. However, due to the inverse cosine function the real part of $n$ has many branches given by

$$n' = \text{Re}\left(\frac{\cos^{-1}\left[(1 - s_{11}^2 + s_{21}^2)/(2 s_{21})\right]}{kd}\right) + \frac{2\pi m}{kd} \qquad (5)$$

where $m$ is an integer. The correct branch can be decided by comparing the results for the different material lengths (different number of unit cells in the propagation direction) [47]. Once $n$ and $z$ are evaluated, effective permittivity and permeability are calculated using $\varepsilon = n/z$ and $\mu = nz$.

Fig. 3 shows the retrieved effective parameters for the four-functional-layer modified fishnet metamaterial. Two DNG bands can easily be observed. The first negative index band exists in the red region of the visible spectrum and extends from 399THz to 462THz with minimum value of $n' = -1.38$. The FOM is 1.66 at the operating frequency of 446THz. The second negative index band exists in the green region of the visible spectrum and extends from 500THz to 587THz having a minimum value of $n' = -1.12$. The FOM is 2.12 at the operating frequency of 544THz. Operating frequency in this report refers to the frequency corresponding to $n'=-1$. The total thickness of four functional layers of the modified fishnet metamaterial is about a factor of 3 smaller than the operating wavelength. The ratio of the free space wavelength to the metamaterial unit cell thickness ($\lambda/d$) in the propagation direction is 13. Negative values for the imaginary parts $(\varepsilon'', \mu'')$ in Fig. 3 are due to the periodicity artifacts [48]. However, passivity of the metamaterial is not violated by the negative values of the imaginary parts of the effective parameters [48, 49].

The small separation between consecutive metal layers of the modified fishnet metamaterial results in strong coupling between consecutive functional layers of the metamaterial. Due to the strong coupling in the modified fishnet metamaterial, a blue-shift in the operating frequency and an increase in the FOM with increased number of functional layers are observed. This observation is consistent with previously reported results in Ref. 47. Fig. 4 shows the real part of the retrieved refractive index for one to four functional layers of the modified fishnet. A blue-shift in the operating frequency can be observed with increased number of layers. However, this shift decreases (thus convergence) with each functional layer.

Care must be taken in assigning the effective parameters obtained using isotropic retrieval to the metamaterial structure. Due to multiple $n'$ branches obtained in the isotropic retrieval, the effective parameters should be compared with other observations. The effective parameters can only be assigned to the metamaterial if all the observations are consistent. For the case of modified fishnet metamaterial, following observations confirm that the effective parameters shown in Fig. 3 correctly describe the metamaterial.

- No discontinuity in any of the $n'$ branches is observed at frequencies below magnetic resonance frequencies. Retrieved parameters provide evidence for magnetic resonances and corresponding negative index bands.
- The $\lambda/d$ ratio is sufficiently large to represent the metamaterial as an effective medium.
- The frequencies at which the peaks are observed in the absorption spectrum (see Fig. 2) align with the relevant resonances in the retrieved effective parameters.
- Simulation results with different number of functional layers of the modified fishnet metamaterial structure show that the retrieved refractive index has a convergent behavior (see Fig. 4).
- Both magnetic resonances corresponding to negative index bands are identified in the magnetic field and current plots (see Fig. 5).

Furthermore, we have verified backward wave propagation in both low and high frequency bands [50]. Figs. 6a and 6b show backward wave propagation inside the 16-layer structure at 495.6THz (orange) and 611.6THz (cyan), respectively. The blue-shift in these frequencies with respect to the dual-band operating frequencies of the 4-layer structure is due to the still converging behavior of the structure (see Fig. 4). The colors correspond to the *x*-component of the magnetic field (i.e., the same component as the incident magnetic field). The direction of the arrows indicates the direction of phase velocity. It is necessary to view the variation of the fields in time to distinguish backward (blue arrows) and forward (red arrows) propagating waves. Media 1 and Media 2 display the time variation of the fields corresponding to Figs. 6a and 6b, respectively. Clearly, the electromagnetic waves inside the structure have negative phase velocity for both frequencies. The lengths of the blue arrows in Fig. 6a correspond to one wavelength inside the structure while the blue arrow in Fig. 6b corresponds to a half-wavelength inside the structure.

These observations confirm that the results presented in Fig. 3 are real and there are two magnetic resonances contributing to negative index bands of the metamaterial. Furthermore, we confirmed, as shown in Table 2, that the bianisotropy introduced by a substrate [51] used in fabrication does not significantly affect the metamaterial behavior.

**Discussion**

We should note that spatial dispersion especially in optical metamaterials may not allow for assigning wave vector independent effective parameters to those inhomogeneous metamaterial structures [52, 53]. Spatial dispersion may cause the retrieved effective parameters strongly depend on the wave vector.

A strong spatial dispersion has been shown in wire media even in the very large wavelength limit [54, 55]. However, for normal incidence (i.e., for the wave propagation in the x-y plane where the electric field is parallel to the wires), a fixed local expression for permittivity is recovered for this specific medium [54]. A local response in the permeability has also been obtained using split-ring-resonators arranged in an orthorhombic lattice under the usual incident field configuration, despite significant spatial dispersion in the perpendicular direction [56]. In contrast with the wire media, based on split ring resonators, it has been shown that the spatially dispersive permeability continuously approaches to the Lorentz local permeability in the long wavelength limit [57]. The same feature in the long wavelength limit has been also observed in Ref. 56. This means that there are also metamaterials which can be represented by effective parameters in the long wavelength limit as opposed to the wire media in Ref. 54. In fact, the fishnet structure is also in this category of metamaterials [58]. It has been concluded in Refs. 54 and 59 that spatial dispersion should be given special consideration to characterize discrete metamaterials as effective media, "at least if arbitrary directions of propagation and/or polarization of the electromagnetic field should be considered in the analysis."

Here, because the structure was intended as one-dimensionally functional negative index metamaterial, the effective parameters were only calculated for normal incident light. Under normal incidence, the plane electromagnetic wave has the same phase over the interaction plane of the structure. Our multiple layer retrieval results (see Fig. 4) show that under normal incidence the structure homogenizes independent of system length. Therefore, physically meaningful dipoles (or meta-atoms) of the homogenous effective medium are created. However, for oblique incidences, these dipoles may be destroyed or become non-local due to strong spatial nonlocality. Therefore, the retrieved effective parameters may lose their usual physical interpretation. In general, especially around the resonance frequencies, where the wavelength inside the structure is short, the interpretation of the retrieved effective parameters becomes cumbersome due to strong spatial dispersion coupled with periodicity artifacts [48].

Convergence of retrieved parameters with an increasing number of metamaterial layers is important because genuine effective parameters have to be length independent [15, 45, 58, 60]. The stronger the damping of the higher order Bloch modes and the less they are excited, the better the effective parameters of a single layer match the effective parameters of the multilayer metamaterial, and the faster the convergence in effective parameters occur [58, 61]. It has been shown that multilayer fishnet structure can be sufficiently represented by bulk effective parameters [55, 58], although the convergence becomes worse for larger angles of incidence due to spatial dispersion [58]. This shows that multilayer fishnet structure (or the modified fishnet structure as it has essentially the same working principle) can be represented by effective parameters under normal or near-normal incident field.

Recently, there have been other homogenization attempts [61-63] to describe the physical meaning of the effective parameters. However, the origin of resonance/antiresonance coupling with these techniques has not been fully understood. Moreover, the characterization of metamaterials with these techniques, especially at optical frequencies, is a big challenge. Therefore, despite abovementioned artifacts, the retrieval procedure has been widely but carefully used by a vast number of researchers due to its straightforwardness.

Although our structure has been designed only for normal incidence and the spatial dispersion may be important for oblique incidence, the dual band feature brings rich opportunities for many interesting applications if the structure is optimized [50, 64]. For example, spatial dispersion in multilayer fishnet structures has been recently studied in Ref. 50 and found that multilayer fishnet structures can be viewed as a general form of indefinite media incorporating not only the electric response but also the magnetic response. The hyperbolic dispersion available in these media can lead to important applications such as subwavelength imaging [65, 66], control and enhancement of spontaneous emission [67], and thermal hyperconductivity [68]. Dual-band operation truly in the visible spectrum, and possibly multiple-band approach, may further extend and enhance these applications as well as underlying physical concepts.

The basic working principle of the modified fishnet is the same as that of a conventional fishnet metamaterial. Therefore, the models developed to analyze the wave propagation through apertures or slits (see for example, Refs. 25-28, 69, 70) can be applied to our modified fishnet structure too. Particularly, we follow the work in Refs. 27 and 28. It has been shown that fishnet metamaterials can support multiple magnetic resonances corresponding to the excitation of different orders of SPP modes. Those multiple magnetic resonances can be used to achieve multiple negative permeability bands, which can in turn contribute to extraordinary backward wave transmission (i.e., negative index) if the standard fishnet structures are modified to increase the filling ratio (hence the effective plasma frequency as shown by the retrieval procedure, see Fig. 3a) without destroying the magnetic resonances. We observed that the extraordinary transmission windows overlap with the regions where the retrieved effective permittivity and effective permeability are simultaneously negative (see Figs. 2 and 3). This result does not only justify the applicability of the retrieval procedure [44-46], but it is also consistent with our previous work [25, 26] as well as with Refs. 27 and 28, where also the retrieval procedure was used.

The low-loss operation is achieved by setting the diluted plasma frequency higher than (but close to) magnetic resonance frequencies. High diluted plasma frequency is obtained by having high metal concentration in a functional layer of the metamaterial. Both magnetic resonances are generated by the excitation of different SPP modes in the metamaterial [27, 71]. Parallel metal layers in the metamaterial provide the inductance and the dielectric spacer between them provides the capacitance to generate net magnetic responses corresponding to two negative index bands. Magnetic field components parallel to the incident magnetic field (i.e., **H**-field in Fig. 1) and corresponding current distributions at absorption peak frequencies for both negative index bands for a single functional layer of the modified fishnet structure are plotted in Fig. 5. Strong magnetic responses can be observed for both resonance frequencies. However, magnetic fields are concentrated at different regions of the metamaterial for the two resonances. The magnetic field is concentrated at the center of the cross-section, for the magnetic resonance at the lower frequency, while for the higher frequency resonance the magnetic field is concentrated at the edges of the cross-section. These strong magnetic resonances oppose the incident magnetic field to produce the negative effective permeability for the metamaterial contributing to both negative indices.

The small separation between consecutive metal layers of the modified fishnet metamaterial results in strong coupling between consecutive functional layers of the metamaterial. The effect of coupling between different functional layers of a fishnet metamaterial has been rigorously studied by Zhou, et al. [47]. They showed that the coupling affects the operating frequency as well as the FOM of the fishnet metamaterial. In general, addition of each functional layer in a fishnet metamaterial increases the overall FOM of the metamaterial and provides a blue-shift in the operating frequency. Furthermore, both the FOM and the operating frequency saturate with increased number of functional layers. The convergence of the operating frequency and the FOM depends on the coupling strength between consecutive functional layers of the metamaterial. In the case of weak coupling, the convergence is very fast and the behavior of multiple functional layers of a fishnet metamaterial is almost the same as that of a single functional layer of the metamaterial. However, in the case of stronger coupling, the convergence is relatively slow and a significant shift in the operating frequency and the FOM can be observed. Similar observations are noted in the modified-fishnet metamaterial (as shown in Fig. 4). Due to the strong coupling, the convergence is very slow and the operating frequency converges after sixteen functional layers of the modified fishnet metamaterial.

Another consequence of the strong coupling is the appearance of hybridization that results in splitting of resonance modes (into symmetric and antisymmetric modes) [47] which can be observed in the retrieved parameters (see Fig. 3) as multiple positive and negative index bands. However, antisymmetric modes (corresponding to negative index bands) have stronger resonance. Furthermore, it must be noted that isotropic retrieval generally ignores the spatial dispersion that produces periodicity artifacts [48]. These periodicity effects result in negative imaginary parts, distortion in the $n'$, and an increase in the FOM [47, 48]. These periodicity artifacts are more prominent in a strongly coupled system.

We should also note that the permittivity of thin metallic layers in our structure should be significantly different than that of a bulk metal in the considered frequency range due to the size-dependent effects such as spatial non-locality [72, 73] and quantum confinement as well as other effects such as increased electron collision rate due to the defects and grain boundaries [74]. The spatial non-locality effect is expected to be relatively small compared to other effects for the metallic thicknesses and inter-particle distances (i.e., distance between metallic layers) in our geometry [72, 73]. The Drude model parameters we have used [43] in the simulations are the fit parameters for the experimental data in Ref. 75 with the exception that the collision frequency is multiplied by three to account for all of the above effects [37, 40, 43].

Another issue is how robust the proposed structure is with respect to the variations in the thicknesses of the individual layers. We analyzed the effect of variation in the material thicknesses for both $MgF_2$ and silver layers. We observed that a slight variation in the thicknesses, on the order of a few nm (i.e., 2-5nm), in particular, an increase in the $MgF_2$ thickness results in a blue-shift (~40THz) in the results, while an increase in the silver thickness results in a slight red-shift (~5THz). In fact, the spectral shift introduced by the variations in the material thicknesses can bring advantages. For example, the present difficulty in depositing a 6nm-thin MgF2 layer cannot only be relaxed by an additional increase in the thickness but also the operating wavelengths can be scaled down further as long as the structure is still described by a homogeneous effective medium and the losses are not substantially high.

**Conclusion**

A modified fishnet metamaterial structure is proposed to achieve a low-loss dual band negative index metamaterial operating in the visible spectrum by adding an additional metal layer of different thickness in a functional layer of a conventional fishnet metamaterial structure. The additional metal layer blue shifts the diluted plasma frequency slightly above the second order magnetic resonance and hence provide two DNG frequency regions, one in the red region and the second in the green region. This is the first dual band negative index metamaterial operating in the visible spectrum that has been reported. The structure operates independent of the polarization of the (normally) incident field. Broadband operation with multiple negative index bands may be possible with the same approach by incorporating additional metal layers and higher order magnetic resonances. The structure can be fabricated by standard microfabrication techniques that have been used to fabricate conventional fishnet metamaterial structures [43, 76].

## Acknowledgments

This work was supported in part by the National Science Foundation under grant ECCS-1202443 and by the Oak Ridge Associated Universities. Authors thank Thomas Koschny at Ames Laboratory for valuable discussions about spatial dispersion and periodicity artifacts in metamaterials. M. I. A. acknowledges for financial support from NED University for his PhD research.
## References

[1]  http://www.wave-scattering.com/negative.html
[2]  A. Schuster, An Introduction to the Theory of Optics, pp. 313-318 (Edward Arnold, London, 1904).
[3]  H. C. Pocklington, "Growth of a wave-group when the group velocity is negative," *Nature* **71**, 607 (1905).
[4]  G. D. Malyuzhinets, "A note on the radiation principle," *Zh. Tekh. Fiz.* **21**, 940 (1951).
[5]  D. V. Sivukhin, "The energy of electromagnetic waves in dispersive media," *Opt. Spetrosk.* 3, 308 (1957).
[6]  V. G. Veselago, "The electrodynamics of substances with simulataneously negative values of e and µ," *Soviet Physics Uspekhi,* vol. 10, pp. 509–514, 1968.
[7]  J. B. Pendry, A. J. Holden, W. J. Stewart, and I. Youngs, "Extremely low frequency plasmons in metallic mesostructures," *Physical Review Letters,* vol. 76, pp. 4773-4776, 1996.
[8]  J. B. Pendry, A. J. Holden, R. D. J., and S. W. J., "Low frequency plasmons in thin-wire structures " *Journal of Physics: Condensed Matter,* vol. 10, pp. 4785–4809, 1998.
[9]  J. B. Pendry, A. J. Holden, D. J. Robbins, and W. J. Stewart, "Magnetism from conductors and enhanced nonlinear phenomena," *Microwave Theory and Techniques, IEEE Transactions on,* vol. 47, pp. 2075-2084, 1999.
[10] D. R. Smith, W. J. Padilla, D. C. Vier, S. C. Nemat-Nasser, and S. Schultz, "Composite medium with simultaneously negative permeability and permittivity," *Physical Review Letters,* vol. 84, pp. 4184-4187, 2000.
[11] R. A. Shelby, D. R. Smith, and S. Schultz, "Experimental verification of a negative index of refraction," *Science,* vol. 292, pp. 77-79, 2001.
[12] C. M. Soukoulis, S. Linden, and M. Wegener, "Negative refractive index at optical wavelengths," *Science,* vol. 315, pp. 47-49, 2007.
[13] C. M. Soukoulis and M. Wegener, "Past achievements and future challenges in the development of three-dimensional photonic metamaterials," *Nature Photonics,* vol. 5, pp. 523-530, 2011.
[14] V. M. Shalaev, "Optical negative-index metamaterials," *Nature Photonics,* vol. 1, pp. 41-48, 2007.
[15] D. Ö. Güney, T. Koschny, M. Kafesaki, and C. M. Soukoulis, "Connected bulk negative index photonic metamaterials," *Optics Letters,* vol. 34, pp. 506-508, 2009.
[16] T. Xu, Y. Zhao, J. Ma, C. Wang, J. Cui, C. Du, and X. Luo, "Sub-diffraction-limited interference photolithography with metamaterials," *Optics Express,* vol. 16, pp. 13579-13584, 2008.
[17] J. B. Pendry, "Negative refraction makes a perfect lens," *Physical Review Letters,* vol. 85, p. 3966, 2000.
17

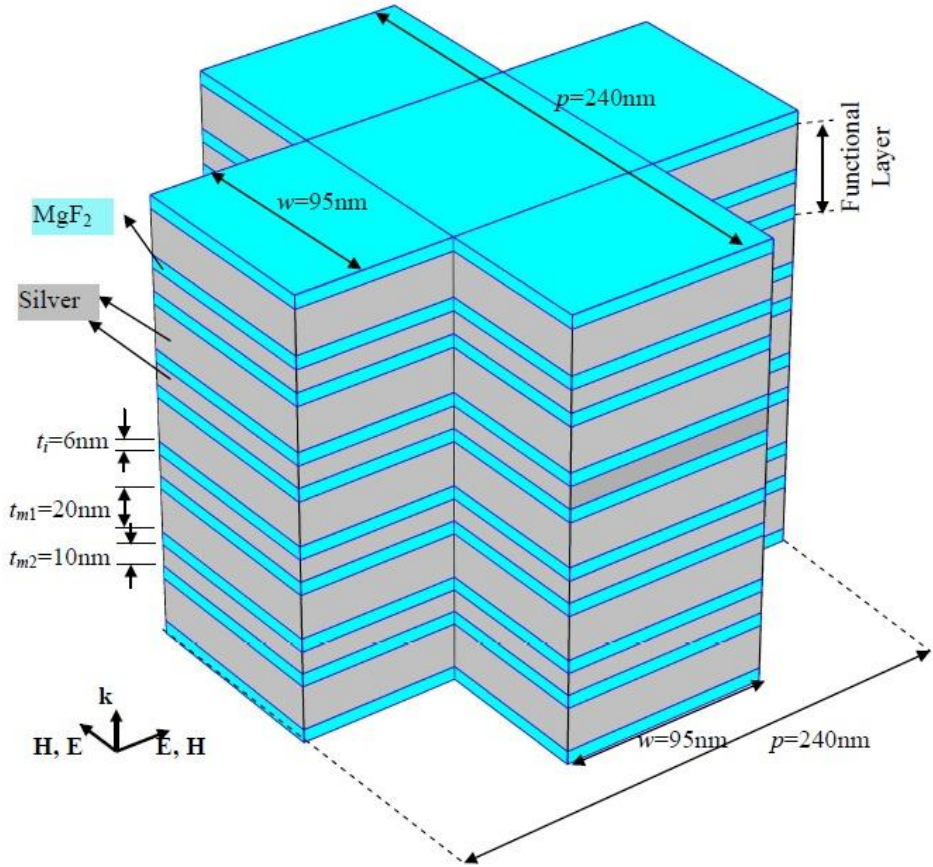

Fig. 1 (Color online) Geometry of the dual-band, polarization-independent, modified fishnet metamaterial for the visible spectrum. Thickness of each dielectric ($MgF_2$) layer is 6nm and that of the alternating metal (silver) layers are 20nm and 10nm. Other dimensions are indicated. The square symmetry of the structure along the lateral directions and subwavelength features ensure the polarization independent behavior.

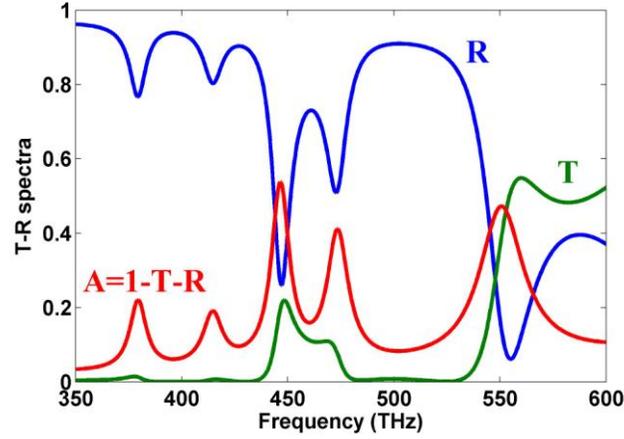

Fig. 2 (Color online) Transmittance (T), reflectance (R), and absorption (A) for the four-functional layer modified fishnet metamaterial. There are two EOT bands; one in the red (around 450THz) and the other in the green region (around 550THz) of the visible spectrum.

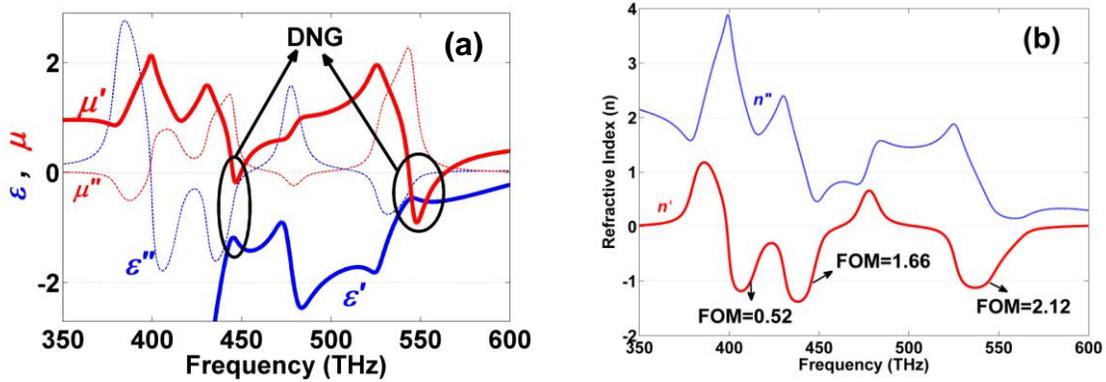

Fig. 3 (Color online) Retrieved effective parameters for the four functional layers of the modified fishnet metamaterial. (*a*) Effective permittivity and permeability of the structure. Regions corresponding to the simultaneously negative $\varepsilon'$ and $\mu'$ are indicated. (*b*) Effective refractive index. There are two DNG operating points with FOMs of 1.66 and 2.12 at 446THz (red) and 544THz (green), respectively.

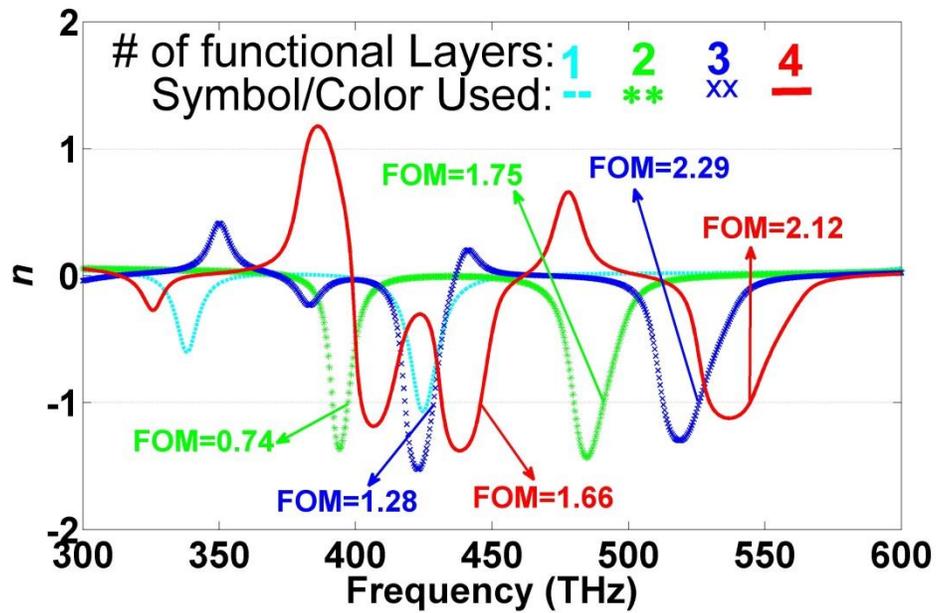

Fig. 4 (Color online) Real part of the effective refractive index for different number of functional layers of the modified fishnet structure. Blue-shift as well as the convergence in operating frequency is apparent with each additional functional layer.

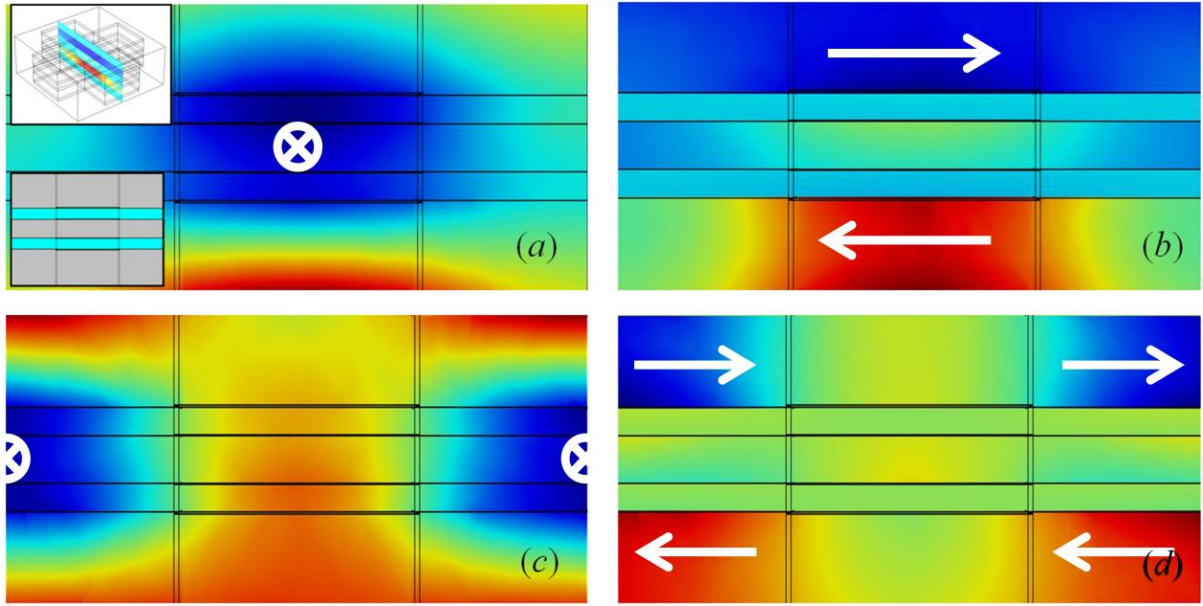

Fig. 5 (Color online) Magnetic field and corresponding current distributions at the central cross-section of the modified fishnet metamaterial shown in the insets where metal gray and cyan colors correspond to metal and dielectric layers, respectively. Other colors in the surface plots indicate the direction and magnitude. The thick white horizontal arrows indicate the direction of major induced currents. These anti-parallel currents result in net magnetic moments that are strong enough to provide negative permeability values. The direction of major induced magnetic fields associated with these anti-parallel currents are indicated by white inward arrows. (a) Magnetic field distribution at the absorption peak of the first negative index band. (b) Current distribution corresponding to (a). (c) Magnetic field distribution at the absorption peak of the second negative index band. (d) Current distribution corresponding to (c).

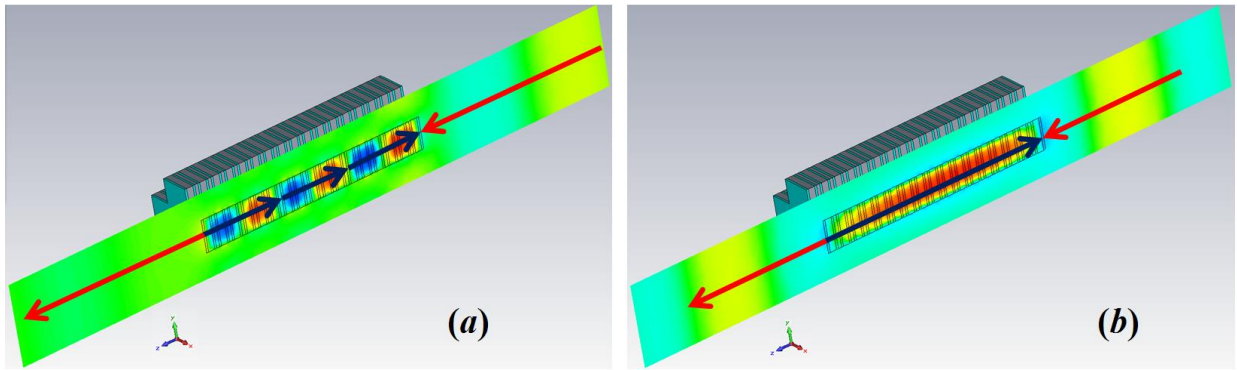

Fig. 6 (Color online) Backward propagation of electromagnetic waves inside dual-band 16-layer modified fishnet structure at (a) 495.6THz (orange) and (b) 611.6THz (cyan). Colors correspond to *x*-component of the magnetic field. Arrows indicate the phase velocity. The time variation of the fields can be viewed in Media 1 and Media 2.

Table 1: Summary of simulation parameters.

| Description | Symbol | Value |
| --- | --- | --- |
| Thickness of each dielectric layer | $t_i$ | 6nm |
| Thicknesses of metal layers | $t_{m1}$, $t_{m2}$ | 10nm, 20nm |
| Lattice constant | $p$ | 240nm |
| Side-length of square aperture | $p - w$ | 145nm |
| Bulk plasma frequency for silver | $f_p$ | 2180THz |
| Collision Frequency of silver | $\gamma_e$ | 13.5THz |
| Permittivity of MgF2 | $\varepsilon_{MgF2}$ | 1.9 |

Table 2: Summary of important observations for modified-fishnet metamaterial for isotropic (without substrate) and bianisotropic (with substrate) cases.

| Number of functional layers | Without substrate | | | With substrate ($n_{Substrate}$=1.52) | | |
| --- | --- | --- | --- | --- | --- | --- |
| | Operating Frequency (THz) | FOM at $n$=–1 | Max (FOM) | Operating Frequency (THz) | FOM at $n$=–1 | Max (FOM) |
| 1 | 339 (IR) | N/A | 0.25 | 334 (IR) | N/A | 0.198 |
| 1 | 426 (red) | 0.7 | 0.71 | 420 (red) | N/A | 0.505 |
| 2 | 397 (IR) | 0.74 | 0.76 | 392 (IR) | 0.59 | 0.61 |
| 2 | 490 (orange) | 1.75 | 1.76 | 485 (orange) | 1.36 | 1.36 |
| 3 | 428 (red) | 1.28 | 1.3 | 426 (red) | 1.10 | 1.2 |
| 3 | 527 (yellow) | 2.3 | 2.6 | 523 (yellow) | 1.95 | 2.1 |
| 4 | 445 (red) | 1.66 | 1.7 | 444 (red) | 1.47 | 1.47 |
| 4 | 544 (green) | 2.12 | 3.15 | 542 (green) | 1.93 | 2.54 |